\documentclass[aps,prb,twocolumn,superscriptaddress]{revtex4-2}
\usepackage{color}
\usepackage{graphicx}
\usepackage{float}
\usepackage{hyperref}
\hypersetup{
    colorlinks=true,
    linkcolor=blue,
    filecolor=blue,      
    urlcolor=blue,
    citecolor=blue
}

\newcommand{\CCI}{CeCoIn$_5$}

\begin{document}

%\preprint{}

\title{Quantum well states in fractured crystals of the heavy fermion material \CCI}
\author{Nicolas Gauthier}
\email{nicolas.gauthier@stanford.edu}
\affiliation{Stanford Institute for Materials and Energy Sciences, SLAC National Accelerator Laboratory, Menlo Park, California 94025, USA}
\affiliation{Geballe Laboratory for Advanced Materials, Departments of Applied Physics and Physics, Stanford University, Stanford, California 94305, USA}
\author{Jonathan A. Sobota}
\affiliation{Stanford Institute for Materials and Energy Sciences, SLAC National Accelerator Laboratory, Menlo Park, California 94025, USA}
\author{Makoto Hashimoto}
\affiliation{Stanford Synchrotron Radiation Lightsource, SLAC National Accelerator Laboratory, Menlo Park, CA 94025, USA}
\author{Heike Pfau}
\affiliation{Stanford Institute for Materials and Energy Sciences, SLAC National Accelerator Laboratory, Menlo Park, California 94025, USA}
\affiliation{Advanced Light Source, Lawrence Berkeley National Laboratory, Berkeley, California 94720, USA}
\author{Dong-Hui Lu}
\affiliation{Stanford Synchrotron Radiation Lightsource, SLAC National Accelerator Laboratory, Menlo Park, CA 94025, USA}
\author{Eric D. Bauer}
\affiliation{MPA-CMMS, Los Alamos National Laboratory, Los Alamos, New Mexico 87545, USA}
\author{Filip Ronning}
\affiliation{Institute for Materials Science, Los Alamos National Laboratory, Los Alamos, New Mexico 87545, USA}
\author{Patrick S. Kirchmann}
\affiliation{Stanford Institute for Materials and Energy Sciences, SLAC National Accelerator Laboratory, Menlo Park, California 94025, USA}
\author{Zhi-Xun Shen}
\affiliation{Stanford Institute for Materials and Energy Sciences, SLAC National Accelerator Laboratory, Menlo Park, California 94025, USA}
\affiliation{Geballe Laboratory for Advanced Materials, Departments of Applied Physics and Physics, Stanford University, Stanford, California 94305, USA}


%\thanks{}

\date{\today}

\begin{abstract}
Quantum well states appear in metallic thin films due to the confinement of the wave function by the film interfaces. Using angle-resolved photoemission spectroscopy, we unexpectedly observe quantum well states in fractured single crystals of \CCI. 
We confirm that confinement occurs by showing that these states' binding energies are photon-energy independent and are well described with a phase accumulation model, commonly applied to quantum well states in thin films. 
%The energy of the quantum well states is independent of photon energy, which confirms their two-dimensional nature. They are well described by a phase accumulation model, commonly applied to quantum well states in thin films.
This indicates that atomically flat thin films can be formed by fracturing hard single crystals. For the two samples studied, our observations are explained by free-standing flakes with thicknesses of 206 and 101~\AA. 
We extend our analysis to extract bulk properties 
%sections of the three-dimensional Fermi surface 
of \CCI. Specifically, we obtain the dispersion of a three-dimensional band near the zone center along in-plane and out-of-plane momenta. 
%The retrieved Fermi surface section corresponds to a hole pocket centered at $\Gamma$.
We establish part of its Fermi surface, which corresponds to a hole pocket centered at $\Gamma$. 
We also reveal a change of its dispersion with temperature, a signature that may be caused by the Kondo hybridization. 

\end{abstract}

%\keywords{}

\maketitle

%%%%%%%%%%%%%%
%%% INTRO %%%
%%%%%%%%%%%%%%
\section{Introduction}

The quantum well is a simple introduction to quantum mechanics that demonstrates the quantization of confined states. Yet, this simple problem is directly relevant in important applications such as diode lasers and quantum dots. The discretization of the energy levels due to confinement occurs in a variety of systems in particle and condensed matter physics. 
For example, in one-dimensional spin chain systems, the interchain interactions generate a potential well that confines the magnetic domain walls~\cite{Coldea2010}.
% interchain interactions in one dimensional spin chains generate a potential well that confines magnetic domain walls~\cite{Coldea2010}.
%in one-dimensional spin chain systems, the interchain interactions generate a potential well that confines the magnetic domain walls~\cite{Coldea2010}.
%the confinement of magnetic domain walls in one-dimensional spin chains~\cite{Coldea2010}. Here, the interchain interactions generate the potential well.
%For example, in condensed matter systems the confinement of magnetic domain walls was observed in one-dimensional spin chains~\cite{Coldea2010}. Here, the interchain interactions generate the potential well. 
In metals, the electronic states can be confined by interfaces, a subject that has been studied thoroughly using photoemission spectroscopy~\cite{Chiang2000}.

In metallic thin films, the electronic states dispersing perpendicular to the film plane are confined by the top and bottom interfaces. The resulting potential boundaries force the formation of quantized wave functions. The study of quantum well states (QWSs) is a well-established and powerful technique that can provide accurate information about single-particle lifetimes and dispersions of three-dimensional bands~\cite{Chiang2000}. The approach was initially applied to elemental metals, such as Cu~\cite{Kawakami1999} and Ag~\cite{Paggel1999a,Paggel1999}, and 
can be used to characterize properties such as electron-phonon coupling. 
%allowed to characterize various properties such as electron-phonon coupling. 
Tuning the quantum well states by varying the film thickness has also been shown to modify the superconducting temperature in Pb~\cite{Guo2004}.
More recent works used QWSs to get further insights on correlated oxides~\cite{Yoshimatsu2011,Santander-Syro2011,Meevasana2011} and characterize topological states in Bi, Sb and Bi$_2$Se$_3$~\cite{Zhang2012a,Ito2016,Zhang2010}.

Here we report the observation of QWSs, signatures typically associated with thin films, via angle-resolved photoemission spectroscopy (ARPES) of fractured single crystals of \CCI. This result is surprising, as a well-defined confinement potential is required to form QWSs.
We demonstrate in section~\ref{secHV} that the states' binding energies are photon-energy independent, as expected for photoemission from two-dimensional (2D) QWSs. In section~\ref{secPAM}, we analyze the observed states using a phase accumulation model, typically employed in thin films. We first obtain the shape of the dispersion along the confined direction without any assumption. The film thickness and quantum numbers of the QWSs are then determined by calibrating our results to previous measurements~\cite{Jang2017}. The results in sections ~\ref{secHV} and \ref{secPAM} confirm that the observed states are QWSs. In sections~\ref{sec3DFS} and ~\ref{secTdep}, we exploit the QWS properties to extract 
bulk properties
%sections of the three-dimensional (3D) Fermi surface 
of \CCI.

\begin{figure*}[ht!]
\includegraphics[scale=1]{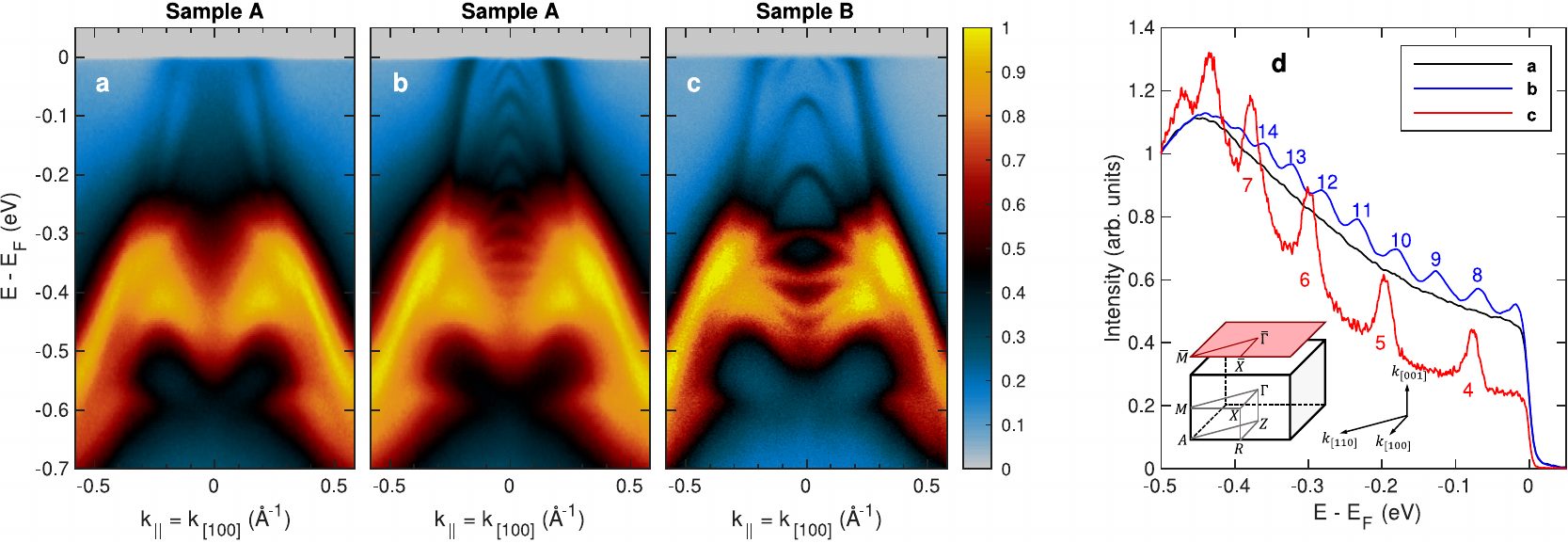}
\caption{Photoemission spectra measured {with $h\nu =25$~eV in $p$-polarization} along the $\bar{\Gamma} \bar{X}$ direction showing the presence of quantum well states (QWSs). Spectra (a) and (b) have been measured on the same sample A but in different regions of its surface. Only the (b) spectrum reveals QWSs near $\bar{\Gamma}$. The spectrum (c) has been measured on sample B and exhibits a different number of QWSs, indicating a different quantization of the QWSs.
(d) Energy distribution curves at $k_\parallel = 0$ from spectra (a) to (c). 
The quantum numbers $n$ determined in section~\ref{secPAM} are shown. 
Inset: Brillouin zone and its surface projection.}
\label{fig1}
\end{figure*}

The material studied, \CCI, is a prototypical heavy fermion with a superconducting state below $T_c = 2.3$~K~\cite{Petrovic2001}. Magnetic interactions, dependent on the degree of itineracy of the $f$-electrons~\cite{Mazzone2019a}, have been proposed to mediate to formation of Cooper pairs in this material~\cite{Petrovic2001} similar to other heavy-fermion materials~\cite{Mathur1998,Monthoux2007}.
Theoretical calculations with itinerant $f$-electrons indicate that the electronic structure of \CCI\ is composed of three bands crossing the Fermi level $E_F$~\cite{Settai2001,Oppeneer2007,Nomoto2014}. These have been investigated with quantum oscillations \cite{Settai2001} and angle-resolved photoemission spectroscopy (ARPES)~\cite{Koitzsch2008,Koitzsch2009,Jia2011,Booth2011,Dudy2013,Koitzsch2013,Jang2017,Chen2017}. 
%One primary interest was to determine the effect of the hybridization of the $f$-electrons with the conduction electrons on the band structure. 
Most experimental observations of the Fermi surface are attributed to the nearly 2D $\alpha$-sheet and the quasi-2D $\beta$-sheet. However, theory also predicts multiple three-dimensional (3D) sheets~~\cite{Settai2001,Oppeneer2007,Nomoto2014,Polyakov2012,Klotz2018}. First, there is a 3D oblate pocket centered at $\Gamma$, which is labeled $\epsilon$ in Ref.~\cite{Settai2001} and $\gamma$ in some ARPES works~\cite{Jang2017,Chen2017}. We will use the $\epsilon$ notation in this work. Second, there is also a 3D ellipsoidal pocket centered at $X$. Finally, there is a more complex 3D sheet around the $ZRA$ plane, labeled $\gamma_Z$ in Ref.~\cite{Jang2017}. A confinement potential along the $c$-axis will mostly affect those 3D bands. Therefore, the QWSs observed in \CCI\ must originate from one of these 3D bands. As explained in section~\ref{sec3DFS}, our results indicate that the QWSs originate from the 3D $\epsilon$ pocket centered at $\Gamma$.

%%%%%%%%%%%%%%
%%% EXP DETAILS %%%
%%%%%%%%%%%%%%
\section{EXPERIMENTAL DETAILS}

Large high quality samples of \CCI\ were grown by flux as reported previously in Ref.~\cite{Petrovic2001}. The typical dimensions of the sample used in the experiment are $2 \times 2 \times 0.4$~mm$^3$. Samples were fixed on a copper post with Silver Epoxy Epo-tek H20E and oriented to expose a (001) surface. A ceramic post was fixed on top of the sample surface with the same silver epoxy and samples were fractured {\it in situ} at temperatures below 20~K. We note that a large transverse force, of the order of a few Newtons, applied on the ceramic post was required to fracture the samples. ARPES measurements were performed at the beamlines 5-4 and 5-2 at Stanford Synchrotron Radiation Lightsource (SSRL). The chamber pressure remained below $4 \times 10^{-11}$~Torr during the measurements. Measurements on sample A, and six other samples, were performed at the beamline 5-2 with a beam spot of $27 \times 43$~$\mu$m$^2$. Linear horizontal (LH) polarized light with a photon energy of 25~eV was used. Complementary measurements were performed with photon energies ranging from $h \nu = 117$ to 127~eV. Measurements on sample B were performed at the beamline 5-4 with a beam spot of $100 \times 200$~$\mu$m$^2$. Measurements were acquired with circular right (CR), linear vertical (LV) and LH polarized light, for photon energies ranging from $h \nu = 7.5$ to 40~eV. Unless otherwise noted explicitly, measurements shown were taken with $h\nu  = 25$~eV in LH polarization. {For both beamline geometries, LV and LH polarizations correspond to $s$-polarized and $p$-polarized light, respectively.}

%%%%%%%%%%%%%%
%%% EXP RESULTS %%%
%%%%%%%%%%%%%%
\section{EXPERIMENTAL RESULTS}

\subsection{Quantum well states in \CCI}

A typical ARPES spectrum of \CCI\ measured with 25~eV photons on sample A along the $\bar{\Gamma} \bar{X}$ direction is presented in Fig.~\ref{fig1}a. Based on the inner potential value of 12 eV~\cite{Dudy2013}, the perpendicular momentum $k_\perp$ is at the Brillouin zone boundary. Therefore, this spectrum corresponds to a cut along the $ZR$ direction and compares well with previous measurements along the same direction with 100~eV photons, also resulting in $k_\perp$ near the Brillouin zone boundary~\cite{Koitzsch2009}. The spectrum is characterized by an intense M-shaped feature between $-0.6$ and $-0.3$~eV, a sharp band extending from it to reach the Fermi level $E_F$, and a continuum filling the region between the M-shaped feature and the Fermi level near $k_\parallel = 0$. 
The observation of a continuum can be explained by a broadening of $k_\perp$ due to final state effects~\cite{Strocov2003}. It must then originate from a 3D band dispersing along the $\Gamma Z$ direction. 

The spectrum presented in Fig.~\ref{fig1}a with its characteristic signatures has been consistently observed on many regions of sample A (see Appendix I) and on other samples. However, from the eight samples we measured, three of them exhibited different spectra in small regions of the surface. Examples of those uncommon spectra in samples A and B are shown in Fig.~\ref{fig1}b-c.
In those spectra, numerous new bands appear near $k_\parallel = 0$ in the energy range between $E_F$ and about $-0.5$~eV. 
We demonstrate in the following that these new bands in \CCI\ are QWSs due to confinement along the $c$-axis. We will show in section~\ref{secHV} that those states are photon-energy independent and in section~~\ref{secPAM} that they are well described by a phase accumulation model. In this model, the different number of bands between samples, as seen in Fig.~\ref{fig1}b-c, is understood from different thicknesses of the confinement region.

\begin{figure}
\includegraphics[scale=1]{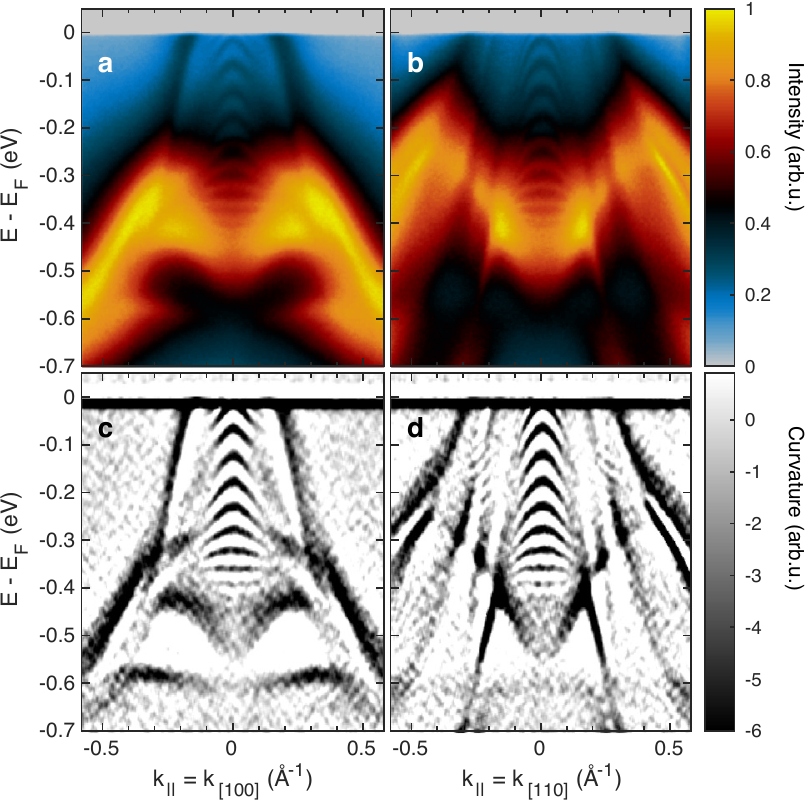}
\caption{Photoemission spectra along the  $\bar{\Gamma} \bar{X}$ direction in (a) and the $\bar{\Gamma} \bar{M}$ direction in (b) measured on sample A. {Measurements were taken using a photon energy of 25~eV in $p$-polarization.} Panels (c) and (d) present the 2D curvature plots associated to panels (a) and (b), respectively. The curvature plots highlight the QWSs near $\bar{\Gamma}$. Some hints of QWSs are also observed around $|k_ {[110]}|=0.25 {\rm \AA}^{-1}$ and $E-E_F=-0.15$~eV in (d).}
\label{fig2}
\end{figure}

The spectrum in Fig.~\ref{fig1}b is reproduced in Fig.~\ref{fig2}a and the QWSs are made more apparent by presenting the 2D curvature of the photoemission intensity~\cite{Zhang2011} in Fig.~\ref{fig2}c. Note that closely spaced levels only appear near $k_\parallel=0$ down to about $-0.45$~eV, in the same energy and momentum range where the continuum is observed in Fig.~\ref{fig1}a. A spectrum along $\bar{\Gamma} \bar{M}$, shown in Fig.~\ref{fig2}b, also exhibits the QWSs near $k_\parallel=0$. Its 2D curvature plot (Fig.~\ref{fig2}d) also reveals less prominent QWSs around $k_\parallel=k_ {[110]}=\pm 0.25\ \text{\AA}^{-1}$ and $-0.15$~eV. 
Only wavevectors in the confined direction, i.e. along $k_\perp$, become quantized.
In our geometry, this corresponds to the $[001]$ direction while $k_\parallel$ is aligned either along the $[100]$ ($\bar{\Gamma} \bar{X}$) or $[110]$ ($\bar{\Gamma} \bar{M}$) direction.
The QWSs that are observed near $\bar{\Gamma}$ in Fig.~\ref{fig2} are associated with a 3D band dispersing along the $\Gamma Z$ direction. This corresponds either to the 3D $\epsilon$-pocket centered at $\Gamma$ or the concave $\gamma_Z$-pocket centered at $Z$. On the other hand, the QWSs identified in Fig.~\ref{fig2}d, around $k_ {[110]}=\pm 0.25\ \text{\AA}^{-1}$, most likely originates from the quasi-2D $\beta$-sheet, which exhibits some dispersion along $k_ {[001]}$.

\subsection{Photon energy dependence}
\label{secHV}

\begin{figure}
\includegraphics[scale=1]{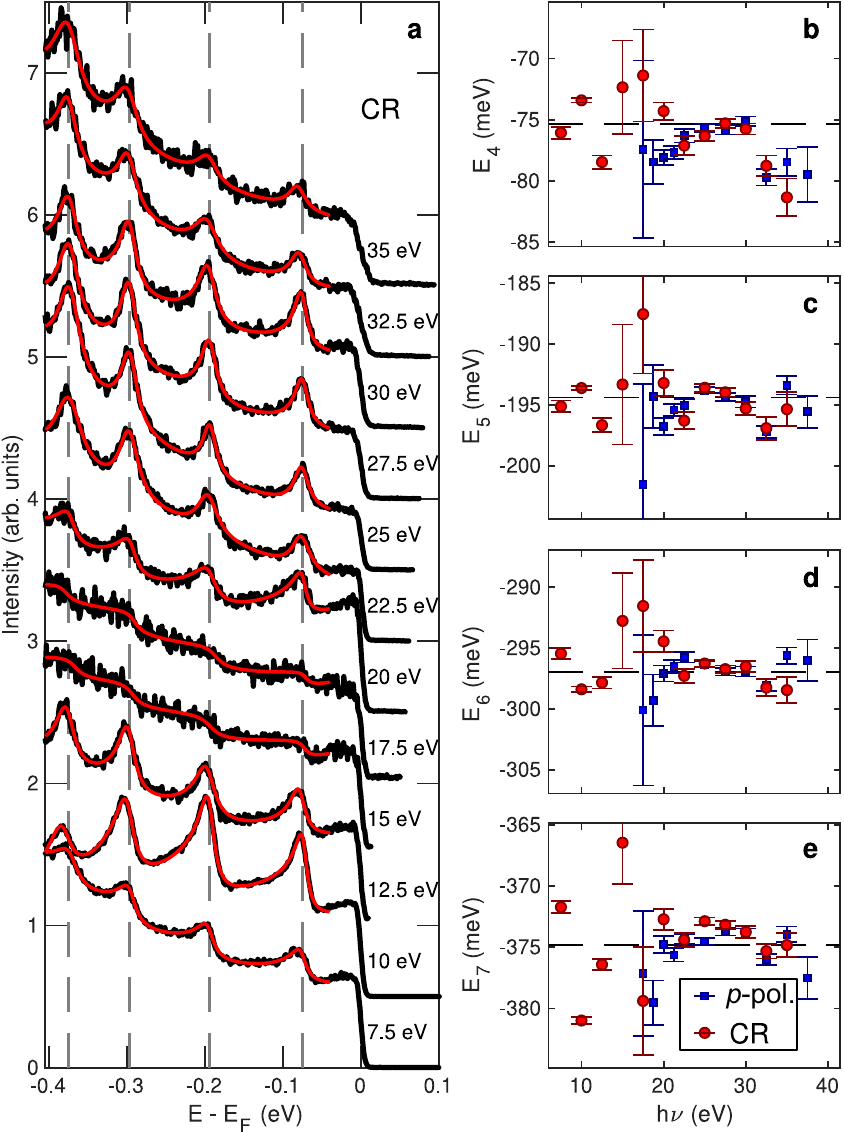}
\caption{Photon energy dependence of the QWSs on sample B. (a) EDCs at $k_\parallel=0$ using circularly polarized light from 7.5 to 35 eV. Red solid lines are fit of four DS lineshapes to the data in black. Vertical gray dashed lines indicate the average energy of each QWS. (b)-(e) Peak positions of the four QWSs nearest to $E_F$ obtained from the DS lineshape fit as function of photon energy.}
\label{fig3}
\end{figure}

Due to their confinement along one direction, QWSs can be understood as 2D states that do not change their binding energy as function of photon energy, in contrast to 3D bulk bands~\cite{Strocov2003}. The binding energies observed in \CCI\ indeed are independent of the photon energy from 7.5 eV up to 37.5 eV, as shown in Fig.~\ref{fig3}. 
%The photon energy dependence reported here was performed on sample B with LH and CR polarized light. 
The photon energy dependence reported here was performed on sample B with {$p$-polarized} and CR polarized light. 
The energy distribution curves (EDCs) measured at $k_\parallel=0$ are qualitatively similar for both polarizations and result in the same binding energy of the QWSs. 

To extract the energies of the QWSs, the EDCs at $k_\parallel=0$ were fitted with 
\begin{equation}
I=A(E) \sum_{n} f_{DS}(E-E_n,\Gamma_n,\alpha) +B(E)
\label{eqDS}
\end{equation}
where $f_{DS}(E-E_n,\Gamma_n,\alpha)$ is a Doniach-\v{S}unji\'{c} (DS) lineshape~\cite{Doniach1970} at energy $E_n$ with a linewidth $\Gamma_n$ and an asymmetry characterized by the parameter $\alpha$, which is caused by the creation of electron-hole pairs. The functions $A(E)$ and $B(E)$ are quadratic polynomials accounting for the peak intensity variation and background, respectively. We fix the asymmetry parameter to be the same for all peaks of an EDC. The resulting fits are presented in Fig.~\ref{fig3}a for the measurements performed with CR polarized light with photon energy from 7.5~eV up to 35~eV.

It is noteworthy that the position of the peak maximum of the DS lineshape $f_{DS}(E-E_n,\Gamma_n,\alpha)$ is different than the excitation energy $E_n$. They only coincide when either $\Gamma_n$ or $\alpha$ tends toward zero. Experimentally, there are small shifts in the position of the peak maximum with photon energy but these can be related to a change of the asymmetry parameter $\alpha$. The peak energies $E_n$ extracted from the fit, presented in Fig.~\ref{fig3}b-e, are effectively independent of the photon energy as expected for QWSs. In the following section, we use the binding energies averaged over all the photon energies to describe the QWSs of sample B at $k_\parallel=0$. For sample A, a comparable photon energy dependence was not performed and the QWS binding energies at $k_\parallel=0$ were instead determined from fitting its EDC in Fig.~\ref{fig1} with Eq.~\ref{eqDS} using $\alpha=0$. 

\subsection{Phase accumulation model}
\label{secPAM}

The binding energies of QWSs in thin films are defined by three important quantities: the film thickness $d$, the band dispersion of the bulk material in the direction perpendicular to the film plane $k_\perp (E)$ and 
%the phase shifts occurring at the boundaries of the confinement potential $\Phi (E)$. 
{the total phase shift $\Phi_{\rm total} (E) = \Phi_{\rm top} (E) + \Phi_{\rm bottom} (E)$ occurring at the top and bottom boundaries of the confinement potential.}
This is expressed mathematically in a phase accumulation model by the Bohr-Sommerfeld quantization rule~\cite{Chiang2000}:
\begin{equation}
2 k_\perp (E) d + \Phi_{\rm total} (E) = 2 \pi n.
\label{eqPAM}
\end{equation}
This equation establishes the conditions to obtain a standing wave in the film. Here, $n$ is the quantum number of the QWS and the next state $n+1$ is reached by adding one node in the standing wave. Typically, measurements are performed on many thin films with different known thicknesses~\cite{Kirchmann2010}. These can be used together with Eq.~\ref{eqPAM} to determine the dispersion $k_\perp (E)$. Here, the nature and size of the confinement potential is a priori unknown. However, the existence of confined states constrains the characteristics of the interfaces. An interface of \CCI\ with another metal would lead to a leakage of the electronic wavefunction into that metal, hindering the formation of QWSs. The top and bottom interfaces must therefore be formed with an insulating medium. The top interface can confidently be assigned to a \CCI-vacuum interface. For simplicity, we assume that the bottom interface is also a \CCI-vacuum interface. While metallic indium inclusions from excess flux can exist in \CCI\ samples, no intergrowth of insulating material is expected. This further supports the assumption of a \CCI-vacuum interface at the bottom{, as vacuum is the only insulating medium present}. This assumption allows to define the total phase shift {$\Phi_{\rm total}(E)$}. The phase shift at a metal-vacuum interface is approximated by:
\begin{equation}
\Phi_{\rm metal-vacuum} (E) = \pi \left( \sqrt{\frac{3.4}{E_V-E} }-1 \right)
\label{eqPhi}
\end{equation}
where $E_V$ is the vacuum level defined by $E_V=E_F+\phi_{\rm metal}$~\cite{McRae1979,McRae1981,Milun2002a}. Here, $\phi_{\rm metal}$ is the metal workfunction. We measured $\phi_{{\rm CeCoIn}_5} = 3.9$~eV with the experimental setup and the method described in Ref.~\cite{Pfau2020}. 

\begin{figure}
\includegraphics[scale=1]{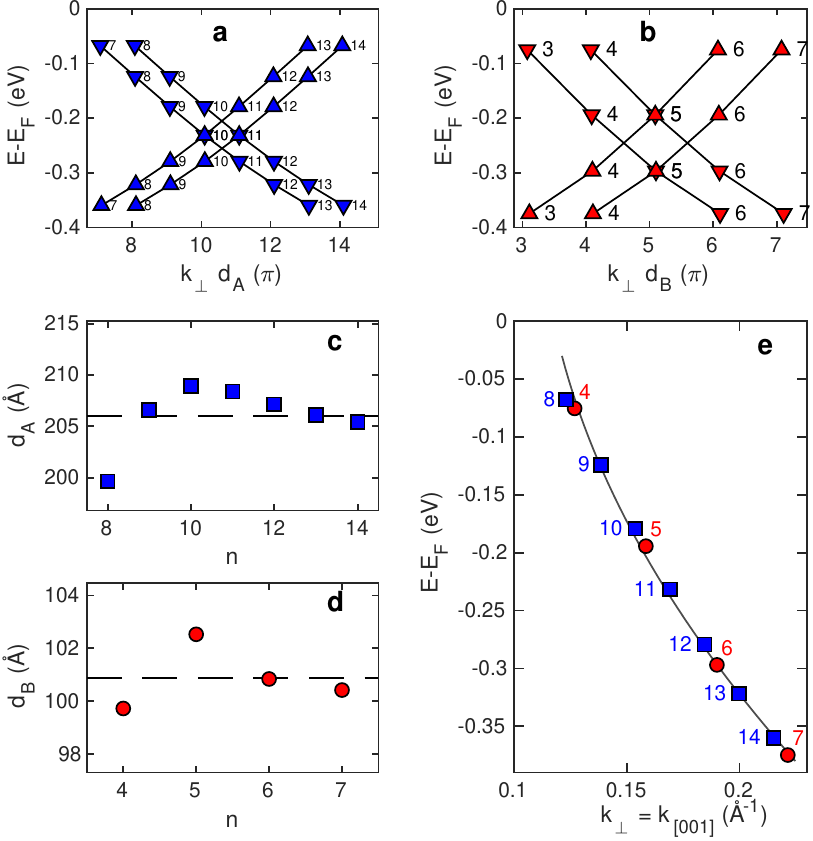}
\caption{(a) Dispersions for sample A, scaled by the thickness $d_A$, along the $\Gamma Z$ direction for four different sets of quantum numbers. A negative (positive) band velocity is obtained for quantum numbers increasing (decreasing) with increasing binding energy. The total phase shift {$\Phi_{\rm total}(E)$} from the metal-vacuum interfaces is included but has a negligible effect on the dispersions. (b) Dispersions for sample B, presented as in panel (a). 
(c)-(d) Calculated thickness $d_A$ and $d_B$ associated to each QWS for samples A and B respectively, for the optimal sets of quantum numbers. The average thickness is indicated by the black dashed line. (e) Dispersion along the $\Gamma Z$ direction based on the QWSs of samples A (blue squares) and B (red circle). The quantum number $n$ of each QWS is indicated. The black line corresponds to the parametrized dispersion used to determine the thicknesses and quantum numbers.}
\label{figThickness}
\end{figure}

The total phase shift ranges from $-0.13\pi$ at {$E-E_F=0$~eV} to $-0.22\pi$ at $-0.4$~eV. This change in the region of interest is small in comparison to the $2\pi$ phase accumulated by each increment of the quantum number. Accordingly, the phase shift has a negligible contribution to the dispersion along $k_\perp$ in Eq.~\ref{eqPAM}. Furthermore, in the infinite box limit {$\Phi_{\rm total}(E) \rightarrow 0$}, the shape of the dispersion is entirely given by the energy spacing between the QWSs. The thickness $d$ and the quantum numbers $n$ can only scale or shift the dispersion on the momentum axis. We illustrate this important point in Fig.~\ref{figThickness}a-b. While we can safely assume that the observed states correspond to a consecutive sequence of quantum numbers, we do not know the offset value or whether this sequence is increasing or decreasing with energy.  
%Using Eq.~\ref{eqPAM}, we show the quantity $k_\perp d$ taken at $k_\parallel=0$ for four different sets of quantum numbers. This quantity corresponds to the dispersion scaled by the unknown film thickness. We observe 
{Using Eqs.~\ref{eqPAM} and \ref{eqPhi}, we evaluate the quantity $k_\perp d$ for the determined energies of the QWSs at $k_\parallel=0$ using four different sets of quantum numbers. The quantity $k_\perp d$ corresponds to the dispersion scaled by the unknown film thickness. The lines in Fig.~\ref{figThickness}a-b connect the data points from one set of quantum numbers and provide a guide to the eye of this scaled dispersion.} We observe 
that choosing increasing or decreasing quantum numbers as function of binding energy changes the sign of the band velocity.
%the band velocity changes sign depending if the quantum numbers are increasing or decreasing with a change in binding energy. 
The dispersion is shifted along the horizontal axis when the offset in the quantum number series is varied. 
%For different sets of increasing (or decreasing) quantum numbers, the dispersion is shifted on the horizontal axis. 
This shows that the shape of the dispersion is determined by the experimental results 
%and not by the 
without any
model parameters.

To pursue our analysis, we rely on data from literature to establish the sign and magnitude of the band velocity as well as its absolute momentum. Cleaving \CCI\ to expose a (100) surface, Jang \textit{et al}. measured the dispersion along $[001]$ (Fig. S3b in Ref.~\cite{Jang2017}), corresponding to the dispersion along $k_\perp$ at $k_\parallel =0$ in our geometry. A well-defined hole-band centered at $\Gamma$ and a weaker diffuse hole-band centered at $Z$ are observed. The former agrees with the expectation of the $\epsilon$ oblate pocket centered at $\Gamma$ from theory~\cite{Elgazzar2004,Oppeneer2007} and de Haas-van Alphen measurements~\cite{Settai2001}. For our analysis, we assume that the QWSs are associated with this hole-band centered at $\Gamma$. Our choice is justified in section~\ref{sec3DFS}. We parametrize the measured dispersion~\cite{Jang2017} between $-0.38$~eV and $-0.03$~eV by a second order polynomial $k_\perp (E)=aE^2 +bE+c$ with $a=0.457$, $b=-0.108$ and $c=0.118$. 

%In the following, we determine a thickness and a set of quantum numbers that scale and shift our results shown in Fig.~\ref{figThickness}a-b to obtain the best agreement with the parametrized dispersion. 
Using this dispersion and Eq.~\ref{eqPAM}, we evaluate the sample thickness for each QWS using a range of consecutive quantum numbers. The objective is to identify a set of quantum numbers for which the calculated thicknesses do not change from one QWS to the next. The set of quantum numbers that lead to the smallest standard deviation of the calculated thicknesses is chosen. The thickness variation for the selected sets of $n$ for samples A and B are presented in Fig.~\ref{figThickness}c-d.  We obtain an average thickness of 206~\AA, or 27.3 unit cells, for sample A and 101~\AA, or 13.4 unit cells, for sample B. The obtained thicknesses are very sensitive to the parametrized dispersion and their absolute values should be taken with caution. Thickness values change by $\sim3\%$ for a shift of the parametrized dispersion by 0.005~\AA$^{-1}$. Larger shifts or changes in the slope can also modify the optimal quantum numbers, leading to more significant changes in the thickness values. Using the average thickness values and the selected sets of quantum numbers, we present the dispersion along $k_\perp$ obtained from the QWSs at $k_\parallel=0$ in Fig.~\ref{figThickness}e. The agreement with the parametrized dispersion, shown by the black solid line, confirms the validity of the approach. 
%We emphasize that although the thickness values should be taken with caution, the shape of retrieved dispersion is robust as illustrated. 

\subsection{3D band structure and Fermi surface}
\label{sec3DFS}

\begin{figure}[h!]
\includegraphics[scale=1]{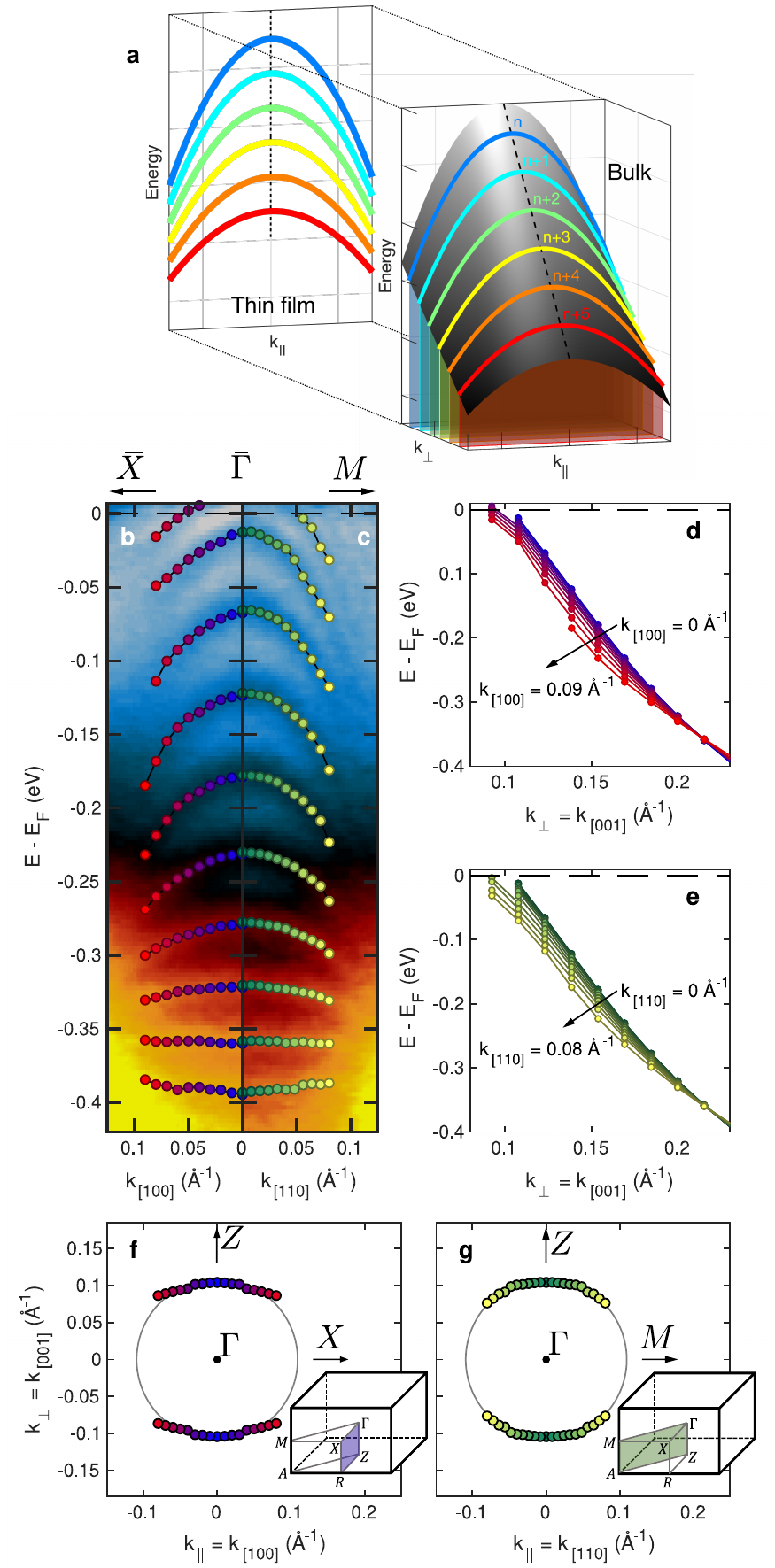}
\vspace{-0.5cm}
\caption{(a) Representation of the effect of confinement on a bulk band dispersing along   $k_\parallel$ and $k_\perp$. Only sections of the bulk dispersion with specific $k_\perp$ values are observed in thin films. (b)-(c) Spectra along the $\bar{\Gamma} \bar{X}$ and $\bar{\Gamma} \bar{M}$ directions measured on sample A showing the QWSs and circles indicating their positions. {The measurements were performed with $h\nu =25$~eV in $p$-polarization.} (d)-(e) Band dispersion along $k_\perp$ for different values of $k_\parallel$. The momentum $k_\parallel$ is along $\bar{\Gamma} \bar{X}$ and $\bar{\Gamma} \bar{M}$ in (d) and (e), respectively. (f)-(g) Retrieved Fermi surface near $\Gamma$ in the $\Gamma Z X$ and $\Gamma Z M$ planes, identified in the Brillouin zone. The circle centered at $\Gamma$ is a guide to the eye.  
%The Fermi surface was determined from a second order polynomial fit of the four data points nearest to $E_F$.
}
\label{figY}
\end{figure}

Fig.~\ref{figY}a illustrates the effect of confinement along $k_\perp$ on a bulk band dispersing along $k_\parallel$ and $k_\perp$. The confinement selects specific $k_\perp$ values of the bulk band, as shown by the colored lines. These sections of the bulk dispersion, or QWSs, are projected on a 2D plane that is measured in thin films. Effectively, 2D spectra $I (E,k_\parallel)$ of QWSs such as the ones in Fig.~\ref{fig2}a-b provide the structure of a 3D object, i.e. the band energy as a function of two different momenta. Up to now, we applied the phase accumulation model only at $k_\parallel =0$ to characterize the dispersion along the $\Gamma Z$ direction, represented by the black dashed line in Fig.~\ref{figY}a. In the following, we extract the complete surface $E(k_\parallel, k_\perp)$ illustrated in gray. Specifically, we determine the band structure $E(k_\parallel = k_{[100]}, k_\perp)$ and $E(k_\parallel = k_{[110]}, k_\perp)$ from the spectra in Fig.~\ref{fig2}a-b. Those spectra are reproduced in Fig.~\ref{figY}b-c, after a division by a Fermi-Dirac distribution convoluted with the energy resolution. The QWS positions, indicated by the circles, were determined by fitting EDCs to Eq.~\ref{eqDS} with $\alpha=0$. Using Eq.~\ref{eqPAM} together with the determined quantum numbers and thickness, the dispersion along $k_\perp$ at each values of $k_\parallel= k_{[100]}$ is calculated and shown in Fig.~\ref{figY}d. These dispersions taken together form the surface $E(k_\parallel= k_{[100]}, k_\perp)$.  The same data treatment is performed for values of $k_\parallel= k_{[110]}$ and is shown in Fig.~\ref{figY}e.

With the established surfaces $E(k_\parallel, k_\perp)$, we are able to determine sections of the Fermi surface associated with this band in two reciprocal lattice planes. For a fixed value of $k_\parallel$, the four data points of $E(k_\parallel, k_\perp)$ closest to $E_F$ are fitted to a second order polynomial. The perpendicular component of the Fermi momentum is determined by the intersection of this polynomial with $E_F$. The associated parallel component of the Fermi momentum corresponds to the fixed value of $k_\parallel$. This procedure is repeated at all values of $k_\parallel$ and the Fermi momenta forming the Fermi surface are traced in Fig.~\ref{figY}f and g for the $\Gamma Z X$ and  $\Gamma Z M$ planes, respectively. In both planes, the Fermi surface follows closely the arc of a sphere centered at $\Gamma$, although there is a small but visible anisotropy. As discussed in section~\ref{secPAM}, we assigned the band forming the QWSs to the $\epsilon$ oblate hole pocket centered at $\Gamma$~\cite{Settai2001,Oppeneer2007}. The theoretical calculations indicate that this pocket is slightly more extended along the $\Gamma X$ direction than along the $\Gamma M$ direction~\cite{Oppeneer2007}. This is in agreement with the general tendency observed for the largest $k_\parallel$ values in Fig.~\ref{figY}f-g and supports our analysis. 

As indicated in section~\ref{secPAM}, a hole band centered at $Z$ is also observed in \CCI~\cite{Jang2017} and it is associated to the $\gamma_Z$-pocket. We performed the same analysis of sections~\ref{secPAM} and \ref{sec3DFS} assuming that this band is responsible for the QWSs, instead of the $\epsilon$-pocket centered at $\Gamma$. This analysis results in a Fermi surface that is inconsistent with the theoretical results. It indicates the formation of an ellipsoidal pocket centered at $Z$ while the theoretical calculations predict a concave pocket with the opposite curvature (see Fig.~S2a in Ref.~\cite{Jang2017}). This disagreement by considering the $\gamma_Z$ band and the agreement by considering the $\epsilon$ band justify the choice made in section~\ref{secPAM}.

\subsection{Temperature dependence}
\label{secTdep}

\begin{figure}
\includegraphics[scale=1]{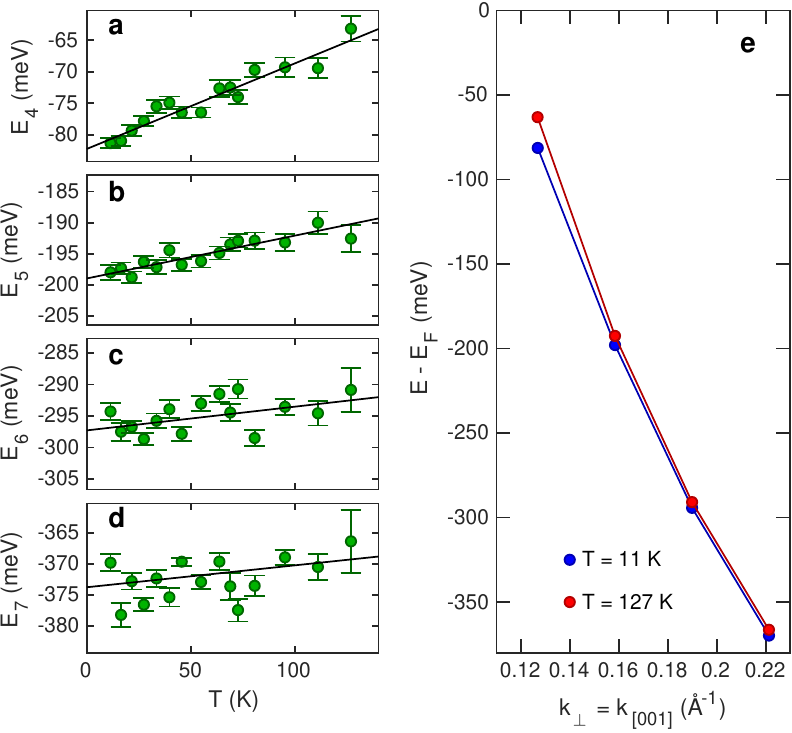}
\caption{Temperature dependence of the QWSs in sample B measured using 20 eV photons with {$p$-polarization}. (a)-(d) Peak position of the QWSs as function of temperature obtained from fitting EDCs at $k_\parallel =0$ to Eq.~\ref{eqDS}. The states nearest to $E_F$ shifts more than the others with temperature. (e)~Dispersion along the $\Gamma Z$ direction obtained from the QWSs at 11 and 127~K. There is small energy shift of the band and also a change in its slope near $E_F$.}
\label{figZ}
\end{figure}

We characterize the temperature dependence of the QWSs for sample B using {$p$-polarized} light with 20 eV photons. By increasing temperature, there is a continuous shift of the QWS binding energy, as shown on Fig.~\ref{figZ}a-d. However, the shift is more pronounced for the states closer to $E_F$. 
The temperature dependence of the QWSs suggests a change in the band structure, which can be caused by a change of the chemical potential as well as a modification of the band dispersion. 

Note that each QWS has a fixed value of $k_\perp$ which is enforced by Eq.~\ref{eqPAM} for a fixed quantum number $n$, assuming that the total phase shift {$\Phi_{\rm total}(E)$} is unchanged. This assumption is reasonable as {$\Phi_{\rm total}(E)$} does not change with temperature for vacuum-metal interfaces and its energy dependence is negligible. The only parameter that is allowed to change is the energy associated to a specific $k_\perp$ value. A rigid band shift caused by a change of the chemical potential would therefore appear as an identical shift of all the QWSs. Experimentally, all the QWSs move towards $E_F$ with increasing temperature, suggesting a change of the chemical potential, but their shift is not identical indicating the contribution from another effect. By reconstructing the $k_\perp$ dispersion at $k_\parallel=0$ at $T=11$~K and 127~K, as shown in Fig.~\ref{figZ}e, we clearly observe a change in the dispersion, concurrently with a small chemical potential shift. This change in dispersion could be related to the formation of the heavy electronic liquid at low temperatures. Previous ARPES measurements on \CCI~\cite{Chen2017,Jang2017} and YbRh$_2$Si$_2$~\cite{Kummer2015,Leuenberger2018} have shown (i) that the Kondo hybridization decreases continuously as function of increasing temperature and (ii) that the spectral signatures of the Kondo effect can be observed far above the Kondo coherence temperature, of 45~K in \CCI. A fading Kondo effect leads to a decrease of the hybridization gap between the renormalized $f$-band and the conduction band. These changes occur close to $E_F$. In \CCI\ one then expects the binding energy of the conduction band to decrease with increasing temperature close to $E_F$ but be temperature-independent further away from $E_F$. This is consistent with our observation. The size of the hybridization gap was found to be linear in temperature up to 200~K for YbRh$_2$Si$_2$~\cite{Leuenberger2018}, which is also consistent with our observation (Fig.~\ref{figZ}a-d).

%%%%%%%%%%%%%%
%%%  DISCUSSION %%%
%%%%%%%%%%%%%%
\section{Discussion}
\label{secDiscussion}

It is surprising that QWSs have not been reported previously in fractured \CCI\ crystals, considering the numerous ARPES works present in the literature~\cite{Koitzsch2008,Koitzsch2009,Jia2011,Booth2011,Dudy2013,Koitzsch2013,Jang2017,Chen2017}. This is most likely because a specific set of conditions is required to observe them. In particular, a small beam spot was important to observe the QWSs in our experiments on \CCI. For example, we used a beam spot of $27 \times 43$~$\mu$m$^2$ at beamline 5-2 and QWSs were only observed in a small region of sample A (see Appendix I). Using a large beam spot or exploring only limited parts of the sample could prevent their observation. 

The characteristics of the light also play a role in the observation of the QWSs. For example, the QWSs were not observed using {$s$-polarized} light (not shown). In {$p$-polarization}, they were hardly visible at 15~eV and 17.5~eV. Previous works have been mostly performed at photon energies higher than 40~eV. In particular, ARPES measurements on \CCI\ are often carried out in the vicinity of the Ce resonance at $h \nu =121$~eV, in search of hybridization effects~\cite{Koitzsch2008,Booth2011,Koitzsch2013,Jang2017,Chen2017}.
We also performed measurements on the QWSs for photon energies of 117 up to 127~eV in {$p$-polarization} on sample A (see Appendix II). We observed that the photoemission cross-section of the QWSs is strongly reduced at these larger photon energies and they only appear in a narrow energy range, from $-0.45$~eV to $-0.2$~eV. Thus, they can easily be overlooked and it could explain why QWSs were not reported earlier.
We point out that the QWS photoemission intensity is unaffected by the Ce-resonance at 121~eV. This is not surprising considering that $f$-spectral weight is not expected in the energy range where the QWSs are observed at this photon energy.

%One of the most intriguing aspect of our results is the observation of QWS in fractured single crystals

% spectroscopic signatures typically associated to thin films, in fractured single crystals. 

One of the most intriguing aspects of our results is the observation, in fractured single crystals, of spectroscopic signatures typically associated to thin films. This suggests that thin film-like structures are formed during the fracturing process. 
%The formation of thin film-like structures by fracturing crystals is one of the most intriguing aspect of our results. 
Our results indicate that we measure structures that are of comparable size to the beam spot ($27 \times 43$~$\mu$m$^2$ at beamline 5-2 and $100 \times 200$~$\mu$m$^2$ at beamline 5-4) and that are atomically flat in order to create well-confined QWSs. Previous STM results showed flat terraces of only tens of nanometers in \CCI\ samples cleaved at room temperature~\cite{Kim2017a}. Thin films grown by molecular beam epitaxy (MBE) also exhibit terraces significantly smaller than 1~$\mu$m~\cite{Haze2018}. 
In contrast, the observation of QWSs requires the presence of two parallel, atomically flat interfaces much larger than those seen in STM on cleaved crystals or MBE-grown films. The coexistence of different thicknesses in the regions probed in our samples is excluded as a single set of QWSs is observed. The coexistence of two thicknesses different by only one unit cell would exhibit QWSs separated by energies similar or larger than the experimental full width at half maximum. % and can be distinguished. 
% sample B would have a shift of about 40 meV, twice the FWHM
% sample A would have a shift of about 20 meV, about the same as the FWHM. The top of the peak would be really flat. 

The fracturing of our \CCI\ crystals was performed at low temperatures ($T\approx 20$~K) by applying a large lateral force on a ceramic post glued to the sample surface. We note that fracturing our samples was rather difficult and required multiple attempts, due to the material hardness, the 3D nature of its structure and the large surface-area-to-thickness ratio of the samples. The unsuccessful attempts could have created cracks, easing the formation of flakes during fracturing. The high load applied is also more likely to induce branching effects known to occur in dynamic fractures~\cite{Bobaru2015}. In that case, a single crack dynamically splits in two cracks, which are expected to follow crystal planes~\cite{Marder2004}. Such a process could create free-standing thin flakes {with parallel atomically flat surfaces. Assuming that such flakes are supported on their edges, vacuum on both sides of the flakes will prevent leakage of the electronic wavefunctions, leading to confinement.} This interpretation suggests that the creation of thin films supporting QWSs might occur more generally in hard crystals when fracturing occurs under high mechanical load. 

The physics of dynamical fractures in materials is rather complex but relevant for their mechanical properties. However, fracturing and its possible consequences are generally not addressed in strongly correlated systems such as \CCI. 
Recently, the effect of fracturing was also considered in the correlated material URu$_2$Si$_2$. In this case, it was shown that a one-dimensional charge density wave can be induced by fracturing at low temperatures~\cite{Herrera2020}. 
This result, together with our observations, highlights that crystal fracturing can reveal unexpected phenomena in strongly correlated systems.

%%%%%%%%%%%%%%
%%% CONCLUSIONS %%%
%%%%%%%%%%%%%%
\section{Conclusions}

In summary, we reported the observation of spectroscopic signatures associated to QWSs in fractured single crystals of \CCI. We confirmed that those signatures are photon-energy independent and are described by a phase accumulation model, as expected for 2D QWSs. The dispersion of the 3D $\epsilon$ hole band along in-plane and out-of-plane momenta was obtained from a detailed analysis of the QWS spectra. From this result, sections of the Fermi surface were extracted and they are in agreement with theory. Finally, a temperature dependence of the QWSs revealed a change of the 3D dispersion that is consistent with a reduction of the Kondo hybridization with increasing temperature.

The observation of QWSs in a fractured single crystal is perhaps the most surprising result of our work. Indeed, a well-defined confinement potential is required to form sharp QWSs. Therefore, our work suggest that fracturing hard crystals at low temperatures can create large atomically flat thin film structures. This interesting observation is however not practical to do a systematic study of QWSs in \CCI. Further work with thin films obtained with controlled growth, which are available for \CCI\ and related materials~\cite{Shishido2010,Mizukami2011,Shimozawa2012,Haze2018}, would be of interest for a deeper investigation on how the Kondo hybridization is affecting the $\epsilon$-band.

The data from this study are available at the Stanford Digital Repository~\cite{Gauthier2020}.

%%%%%%%%%%%%%%
%%% acknowledgments %%%
%%%%%%%%%%%%%%
\begin{acknowledgments}
This work was supported by the Department of Energy, Office of Basic Energy Sciences.
N.G. acknowledges support from the Swiss National Science Foundation (fellowship no. P2EZP2 178542).
H.P. acknowledges support from the German Science Foundation (DFG) under reference PF 947/1-1 and from the Advance Light Source funded through U.S. Department of Energy, Office of Science. 
The Stanford Synchrotron Radiation Lightsource, SLAC National Accelerator Laboratory, is supported by the U.S. Department of Energy, Office of Science, Office of Basic Energy Sciences.
Work at Los Alamos was performed under the auspices of the U.S. DOE, Basic Energy Sciences, Division of Materials Sciences and Engineering.
N.G. is thankful to D.G. Mazzone for fruitful discussions.
\end{acknowledgments}

%%%%%%%%%%%%%%
%%% APPENDIX %%%
%%%%%%%%%%%%%%
%\appendix
\vspace*{0.2 cm}
\section*{Appendix I}

\begin{figure}[htb!]
%\vspace{0.5 cm}
\includegraphics[scale=1]{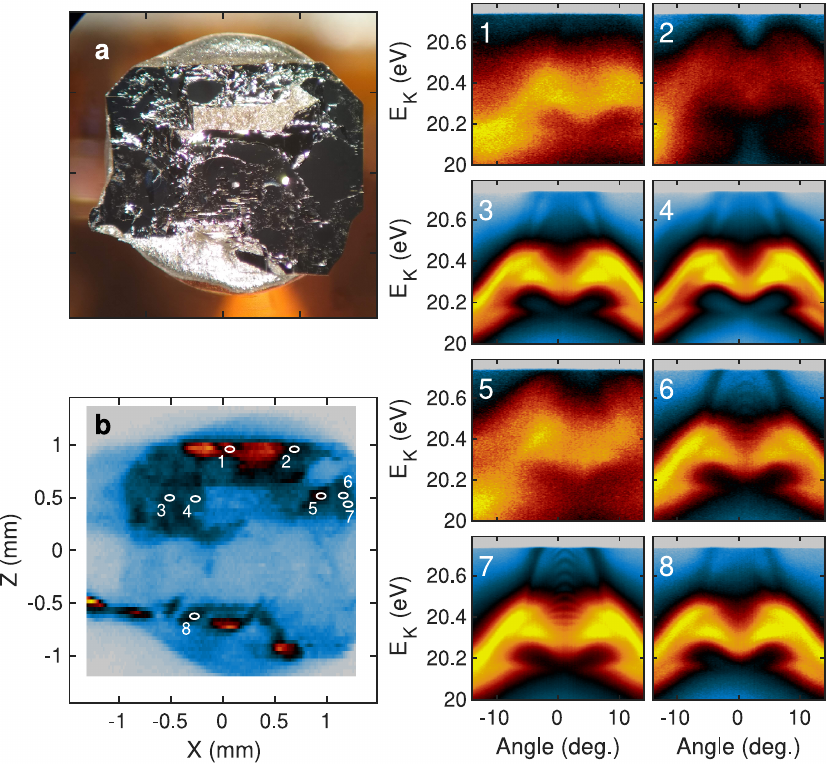}
\vspace{-0.0 cm}	
\caption{(a)~Picture of the sample A after fracturing. (b)~Photoemission intensity map of the sample A. {The intensity was integrated over the full angular range ($\pm 15^\circ$) and 20.51~eV~$< E_K < 20.86$~eV.} White ellipses show different measurements positions (1 to 8). The size of the ellipse represents the beam spot size. On the right, spectra collected at the eight measurement positions are presented. Measurements were performed with 25~eV photon with {$p$-polarization}, along the $\bar{\Gamma} \bar{X}$ direction.}
\label{figApp}
\end{figure}

Spectra obtained at different regions of the sample A are presented in Fig.~\ref{figApp}. Regions 1, 2 and 5 have diffuse spectra, while sharp bands are observed in regions 3, 4, 6, 7 and 8. From these spectra, only regions 6 and 7 exhibit QWSs. Regions 3,4 and 8 instead have diffuse intensity near $\bar{\Gamma}$.

\section*{Appendix II}

\begin{figure}[htb!]
\includegraphics[scale=1]{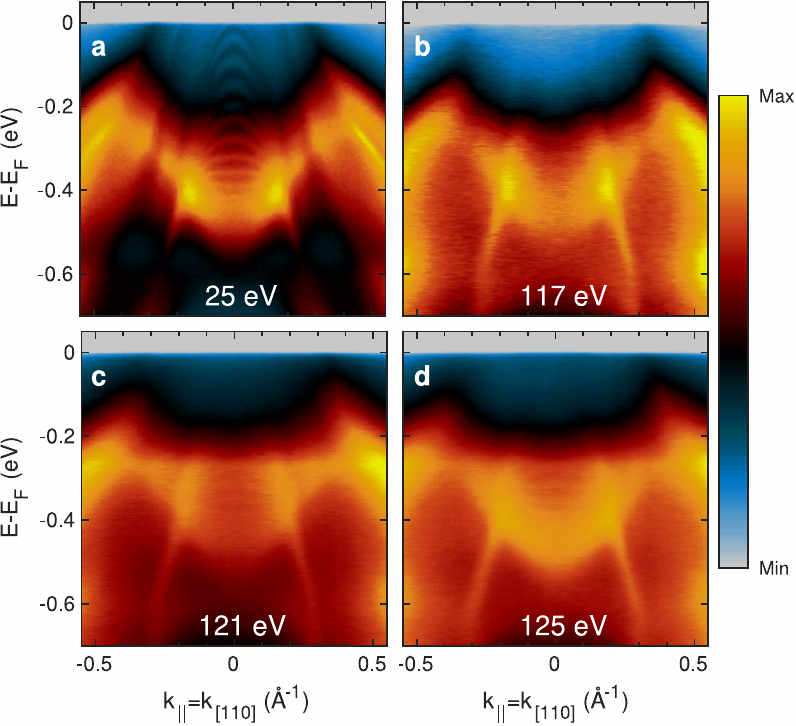}
\vspace{-0.0 cm}
\caption{(a)-(d) Photoemission spectra along the $\bar{\Gamma}\bar{M}$ direction measured on sample A with photon energies of 25, 117, 121 and 125~eV using {$p$-polarization}.}
\label{figApp2}
\end{figure}

In the main text, the measurements of QWSs for photon energies below 40~eV are reported. Here, we present measurements at photon energies around the Ce-resonance of 121~eV on sample A along the $\bar{\Gamma}\bar{M}$ direction. The QWSs are also observed in this energy range, as shown in Fig.~\ref{figApp2}b-d. The contrast between the QWSs and the background is however strongly reduced, in comparison to the spectrum measured at 25~eV (Fig.~\ref{figApp2}a). 

\bibliography{refQW1}

\end{document}